\documentclass{ws-ijmpa}
\usepackage[super,compress]{cite}
\usepackage{graphicx}
\usepackage{float}
\restylefloat{table}
\usepackage{slashed}
\usepackage{mathtools}
\usepackage{dcolumn}
\usepackage{longtable}
\RequirePackage{graphicx}
\RequirePackage{mathptmx}      
\RequirePackage{flushend}
\RequirePackage[numbers,sort&compress]{natbib}
\RequirePackage[colorlinks,citecolor=blue,urlcolor=blue,linkcolor=blue]{hyperref}
\DeclarePairedDelimiter\bra{\langle}{\rvert}
\DeclarePairedDelimiter\ket{\lvert}{\rangle}
\DeclarePairedDelimiterX\braket[2]{\langle}{\rangle}{#1 \delimsize\vert #2}
\DeclarePairedDelimiterX\inner[2]{\langle}{\rangle}{#1,#2}

\begin{document}

%
\catchline{}{}{}{}{}
%

\title{Mass spectra and decays of ground and orbitally excited $c\bar{b}$ states in non relativistic quark model
}

\author{Antony Prakash Monteiro
}

\address{P. G. Department of Physics, St Philomena college
              Darbe, Puttur  574 202, India\\
aprakashmonteiro@gmail.com}

\author{Manjunath Bhat}

\address{P. G. Department of Physics, St Philomena college
              Darbe, Puttur  574 202, India\\
manjunathbhat61@yahoo.in}

\author{ K. B. Vijaya Kumar}

\address{Department of Physics, Mangalore University,
Mangalagangothri P.O., Mangalore - 574199, India\\
kbvijayakumar@yahoo.com}
\maketitle

\begin{history}
\received{Day Month Year}
\revised{Day Month Year}
\end{history}

\begin{abstract}
The complete spectrum of $c\bar{b}$ states is obtained in a phenomenological non relativistic quark model(NRQM), which consists of a confinement potential and one gluon exchange potential (OGEP) as effective quark - antiquark potential. We make predictions for the radiative decay (E1 and M1) widths and weak decay widths of $c\bar{b}$ states in the framework of NRQM formalism.

\keywords{Mesons;Phenomenological quark models;Non relativistic quark models;Leptonic;semileptonic;radiative decays of mesons}
\end{abstract}

\ccode{PACS: 14.40.-n;12.3.-x,12.39.-Jh,13.20.-v}


\section{INTRODUCTION}

\label{sec:intro}

The $B_c$ state is the only bound system that consists of two heavy quarks of different flavours that offers a sound laboratory opportunity to observe both QCD and weak interaction. The ground state  $B_c(1S)$ was first observed in the CDF and DO experiments (Tevatron) in two decay modes $\colon$ $B_c\rightarrow J/\psi l\nu$ and $B_c\rightarrow J/\psi \pi$ \cite{FA98,AA06,AA706,TAA07,KA,VM08,VM09}, also observed by LHC experiments in the various decay modes of $c\bar{b}$ states such as $\colon$ $B_c\rightarrow J/\psi\pi$(LHCb, CMS, ATLAS), $B_c\rightarrow J/\psi\pi\pi\pi$ (LHCb and CMS),~~~ $B^+_c\rightarrow\psi(2S)\pi^+$, $B^+_c\rightarrow J/\psi D^+_s$, $B^+_c\rightarrow J/\psi D^{*+}_s$, $B^+_c\rightarrow J/\psi K^+$, $B^+_c\rightarrow J/\psi 3\pi^+2\pi^-$ and $B^+_c\rightarrow J/\psi K^+K^-\pi^+\pi^-$ (LHCb) \cite{LH12,LH13,LH131,LH132,LH133,JH14}. Recently the ATLAS collaboration at the LHC has observed radially excited $c\bar{b}$ state (i.e, $B_c(2S)$) through the decay channel $B^\pm_c(2S)\rightarrow B^\pm_c(1S)\pi^+\pi^-$ \cite{AG14}. $B_c$ mesons are predicted by the quark model to be members of the $J^P=0^-$ pseudo scalar ground state multiplet \cite{EC94}. The vector $B^*_c(1S)$ meson is the triplet state of $B_c(1S)$ which has not been observed by means of experiments to date. \\

The discovery of $B_c$ state has made the families of well  investigated charmonium and bottomonium complete and has given a new insight into the study of heavy quark dynamics. Since the charmed bottom meson $c\bar{b}$ is an intermediate state of the $c\bar{c}$ mesons and $b\bar{b}$ mesons, its analysis could give detailed information on the balance between the perturbative and non perturbative effects. The investigation of masses of $c\bar{b}$ states gives us an opportunity to obtain information on the nature of the strong interaction thereby it throws up an interesting issue and a tantalising problem. Since the flavour asymmetry of $c\bar{b}$ state unlike in symmetric quarkonium, forbids the annihilation of $c\bar{b}$ state into gluons, the ground state of $c\bar{b}$ state below the BD threshold, can only decay through weak interaction that provides an ideal platform to study weak decays and provides new methods for calculating the CKM matrix. Some  decay channels of $c\bar{b}$ states show that bound state effects are significant in $c\bar{b}$ decays. The lighter c quark has a greater decay rate (65$\%$) than the heavier b quark (25$\%$).  The pseudo scalar $c\bar{b}$ state decays only weakly. The $c\bar{b}$   is unique in that either one
of its quarks can decay, leaving the other as a spectator or both the quarks may involve in its weak decay. Thus $B_c$ meson can serve as a great laboratory for QCD sum rules, Heavy Quark Effective Theory, lattice QCD and potential models. \\

   This work uses NRQM formalism to study both mass spectra and decay properties of $c\bar{b}$ states. The NRQM formalism is found to provide systematic treatment of the perturbative and non perturbative components of QCD at the hadronic scale \cite{BC81,SN85,GS76,SG85,CP00,BG97}. The masses of the $c\bar{b}$ spectrum can be predicted using NRQM whose parameters are tuned to reproduce the spectra of the observed charmonium and bottomonium states \cite{SG88,YY92,VA95,EC94,MG00,WJ91}.   There are a good number of theoretical models that study leptonic, semileptonic and hadronic decay channels of $c\bar{b}$ state \cite{WY07,EH06,JF08,DE03,PF02,VV00,SSG95,AKL10,AVL12,AVL}.\\
     
     In the last two decades, the lattice QCD Monte Carlo calculations  have emerged as a reliable non-perturbative method to study hadron spectra. For $q\bar{q}$ systems it has been shown unambiguously that the ground state potential is $V_{q\bar{q}}=-\frac{A_{q\bar{q}}}{r}+\sigma_{q\bar{q}}r+C_{q\bar{q}}$, with inter quark distance r \cite{TT01,TT02,GSB92}, which is consistent with  the standard NRQM potential of Coulombic + OGEP + linear confinement\cite{YN74}. Also, from lattice QCD, the effect of the gluonic excitation  in the three quark system has been investigated.  It  has been shown that for low-lying hadrons with excitation energy below 1 GeV, the effect of the gluonic excitations is negligible and hence quark degrees plays a dominant role in low-lying hadrons and hence resolves the absence of gluonic excitation modes in low-lying hadron spectra\cite{TT04}. Also, the static three-quark potential has been studied in detail using SU(3) lattice QCD. The detailed analyses lattice QCD data of the 3q potential support the Y-ansatz\cite{TT02}.\\

In the recent work on lattice QCD, an investigation to estimate the masses of B,$B_s$ and $B_c$ has been carried out. The work takes into account of the effect of the u, d and s sea quarks. The highly improved staggered quark (HISQ) action for u/d,s  and c quarks and non-relativistic QCD is used for b quarks. Using HISQ formalism, the mass difference and decay constant ratios between B,Bs and Bc mesons. The mass obtained for $B_c$ mesons is 6280(10) GeV.  which is a significant improvement  on an earlier value of Bc mesons obtained from lattice QCD which is a strong test of lattice QCD to the B physics\cite{EB11}. \\
 
      Our basic aim is to develop a consistent model which could reproduce both the spectra and the decay widths with the same set of parameters. The parameters in our  work, are fixed from the mass spectroscopy and the same set of a parameters have been used to obtain the decay widths.  From our analysis we infer that our model has the right prediction both for the mass spectrum and decay widths with the same set of parameters. It should be noted that obtaining the mass spectrum alone in accordance with the experimental results doesnot guarantee the validity of a model for describing hadronic interactions. Different potentials can reproduce the same spectra. Hence, in a given model, one must be able to calculate other observables like the decay constants, leptonic decay widths, the radiative decay widths, etc. Heavy quarkonium  decays provide a deeper insight on the exact nature of the inter quark forces and decay mechanisms. For example, the leptonic decay widths are a probe of the quarkonium system provide important information complementary to level spacings.  In the existing quark models,  The OGEP has its origin in the exchange of a single gluon which belongs to an octet representation of the $SU(3)_c$. The OGEP is obtained from the QCD Lagrangian in the non-relativistic by retaining terms to the order of $1/c^2$. This procedure is similar to the derivation of the Fermi-Breit interaction in quantum electrodynamics\cite{BEL82}. In deriving OGEP,the gluon propagtaors used are similar to the free photon propagators used in obtaining  Ferm-Breit interaction in QED. Since the confinement of color means confinement of quarks as well as gluons, the confined dynamics of gluons should play a decisive role in determining the spectroscopy of the mesonic states. But, Fermi-Breit interaction which gives rise to  the splitting for singlet and triplet states are treated as perturbation. But, the OGEP is attractive for singlet states and repulsive for triplet states, hence, naïve perturbative treatment of OGEP is incorrect. This leads to further renormalization of strength of interaction for a better fit\cite{MO91}. Also, the most prominent flaw of non-relativistic potential models is the neglect of gluon dynamics\cite{DG75,OW78,DB87,BC81}. Hence, it is required to obtain the mass spectrum by  diagonalizing the Hamiltonian matrix.  \\
   
  The paper is organized in 4 sections. In sec.~\ref{sec:TB} we briefly review the theoretical background for non relativistic model, the description of radiative and weak decay widths. In sec.~\ref{sec:RD} we discuss the results and the conclusions are drawn in sec.~\ref{sec:C} with a comparison to other models. 
\section{THEORETICAL BACKGROUND}
\label{sec:TB}
\subsection{The Hamiltonian}
Essentially, in all phenomenological non relativistic QCD based quark models, the Hamiltonian for the quark system consists of the kinetic energy, the two-body confinement potential and OGEP. In most of these works, it is assumed that the principal binding forces of hadrons are the long range quark confining forces. These should be independent of quark spins and quark masses, depending just on the spatial separations of the constituent quarks. In addition there exist short range forces depending on the quark spins and masses. The effective short range force stems from the one gluon exchange mechanism. The exchange of gluons can provide binding between quarks in a hadron.
The Hamiltonian employed in our model  ~\cite {VH04,BS08, VB09} includes kinetic energy part, confinement potential and one gluon potential (OGEP) \cite{DG75}.
\begin{equation}
H=K+V_{CONF}+V_{OGEP}
\end{equation}
 The kinetic energy part (K)  is the sum of the kinetic energies including the rest mass
minus the kinetic energy of the center of mass motion (CM) of the
total system, i.e., 

\begin{eqnarray}
K & = & \left[\sum_{i=1}^2 M_{i} + \frac{P^{2}_{i}}{2 M_{i}} \right]
- K_\mathrm{cm}, \label{eq:K}
\end{eqnarray}

with $M_{i}$   and $P_{i}$ as the mass and momentum of the $i{\rm
th}$ quark, respectively. $K_{CM}$ is the kinetic energy of the centre of mass motion.

The confinement potential must come ultimately from a non-perturbative treatment of QCD, whereas the residual interaction OGEP is based on perturbation theory. In phenomenological quark models the confinement potential is assumed to be harmonic oscillator potential ($V\sim r^2$) or logarithmic potential ($V\sim ln(r)$) or linear potential ($V\sim r$). For our model we have chosen the linear potential which represents the non perturbative effect of QCD that confines quarks within the color singlet system ~\cite{BS08, VB09}.
\begin{equation}
V_{CONF}(\vec{r}_{ij})= -a_{c}r_{ij}\vec{\lambda}_{i}\cdot
\vec{\lambda}_{j} \label{eq:V-conf}
\end{equation}

where $a_{c}$ is the confinement strength and $\lambda_{i}$ and $\lambda_{j}$ are
the generators of the color SU(3) group for the $ i{\rm th}$ and $ j{\rm th}$ quarks.\\

It should be noted that the two body confinement potential\cite{KS89}, has symmetric and antisymmetric terms. 
\begin{equation}
V^{conf}_{so}=\sum_{i>j}-\frac{1}{4m^2r_{ij}}\frac{dV^{conf}}{dr_{ij}}\{(r_{ij}\times p_{ij})\cdot(\sigma_i+\sigma_j)+[r_{ij}\times(p_i+p_j)/2]\cdot(\sigma_i-\sigma_j)\}
\end{equation}
If one includes both the symmetric and antisymmetric terms the two terms of $V^{conf}_{ij}$ cancel each other and give an almost vanishing contribution to the single baryon/meson spectra. Also, it should be noted that the symmetric term of the $V^{conf}_{ij}$ has opposite sign to the one in $V^{OGEP}_{ij}$ and hence cancels the contribution from the symmetric term of $V^{conf}_{ij}$ in single baryon/meson system. The antisymetric term is Galilei non-invariant. 
It should be noted that the spin orbit term in $V^{conf}_{ij}$ comes from the relativistic effects. The above two body potential is having problems with the long range attractive color Van der Waals force\cite{RS78,PR78,SM79,OW81} when it is used for the two hadron system. Since the attraction between color singlet hadrons comes from the virtual excitation of the color octet dipole state of each  hadron, this problem is evaded by restricting the model space to describe hadrons. Also, if the two body confinement mechanism is extended to two nucleon system, it essentially gives a zero contribution to the spin-orbit force of the N-N potential\cite{FM88}. Since the mechanism which we take for confinement, namely two body confinement, is purely phenomenological it is advisable to leave out the spin-orbit term due to confinement.\\

We consider a purely linear confinement potential in our calculation. The spin-orbit splittings calculated in our model\cite{APM11} and in \cite{WL91} for heavy quark system such as bottomonium and charmonium suggests a scalar confinement. W. Lucha {\textit{et al.}\cite{WL91}, Bhagyesh {\textit{et al.}\cite{BG12,BKB11} and Bhagyesh and K. B. Vijaya Kumar \cite{BKB13} used a mixture of scalar and vector confinement potential to explain the mass spectra and decays of heavy quarkonia. In our calculation we use a one gluon exchange potential and a purely linear confinemnt potential. The one gluon exchange potential is of purely vector nature and the confinement potential is of purely scalar nature. The combination $V(r)=V_{OGEP}+V_{CONF}$ is a mixture of  Lorentz scalar and Lorentz vector nature.\\

The one gluon exchange potential is given by
\begin{equation}
V_{OGEP}=-\frac{4}{3}\frac{\alpha_s}{r}+V_{SD}(r)
\end{equation} 

where the spin dependent potential $V_{SD}$ is introduced as an additional term to the potential to take into the account the spin-orbit and spin-spin interactions, causing the
splitting of the $nL$ levels (n is the principal quantum number, L is the orbital momentum), so it has the form \cite{EF81,GD84,VA95}
\begin{equation}
\begin{split}
V_{SD}(r)=&\left(\frac{L\cdot S_c}{2m^2_c}+\frac{L\cdot S_b}{2m^2_b}\right)\left(-\frac{dV(r)}{rdr}+\frac{8}{3}\alpha_s\frac{1}{r^3}\right)
+\frac{4}{3}\alpha_s\frac{1}{m_cm_b}\frac{L\cdot S}{r^3}+\frac{4}{3}\alpha_s\frac{2}{3m_cm_b}S_c\cdot S_b 4\pi\delta(r)\\
&+\frac{4}{3}\alpha_s\frac{1}{3m_cm_b}\left[3(S_c\cdot n)(S_b\cdot n)-S_c\cdot S_b\right]\frac{1}{r^3}
\end{split}
\end{equation} 
The central part of the two-body potential due to OGEP is ~\cite{DG75},
\begin{equation}
V_{OGEP}(\vec{r}_{ij})=\frac{\alpha_s}{4}\vec{\lambda}_i\cdot\vec{\lambda}_j\left[\frac{1}{r_{ij}}-\frac{\pi}{M_iM_j}\left(\frac{M_i}{M_j}+\frac{M_j}{M_i}+\frac{2}{3}\vec{\sigma}_i\cdot\vec{\sigma}_j\right)\delta(\vec{r}_{ij})\right]
\label{eq:V-ogep}
\end{equation}
where the first term represents the residual Coulomb energy and the
second term is the chromo-magnetic interaction leading to the
hyperfine splitting. $\sigma_{i}$ is the Pauli spin operator and
$\alpha_{s}$ is the quark-gluon coupling constant.

The non$-$central part of OGEP has two terms, namely the spin$-$orbit interaction $V^{SO}_{OGEP}(\vec{r})$ and tensor  term $V^{ten}_{OGEP}(\vec{r})$. The spin-orbit interaction of OGEP is given by,
\begin{equation}
V^{SO}_{OGEP}(\vec{r})=-\frac{\alpha_s}{4}{\bf \lambda_i\cdot\lambda_j}\left[\frac{3}{8M_iM_j}\frac{1}{r^3}(\vec{r}\times\vec{p})\cdot({\bf \sigma_i}+{\bf \sigma}_j)\right]
\end{equation}
where the relative angular momentum is defined as usual in terms of relative position $\vec{r}$ and the relative momentum $\vec{p}$. Unlike the tensor force, the spin$-$orbit force does not mix states of different $\vec{L}$, since $L^2$ commutes with $\vec{L}\cdot\vec{S}$, $\vec{L}$ is still a constant of motion, but $L_z$ is not.

We use the following tensor term \cite{BA89, VB95}
\begin{equation}
V^{ten}_{OGEP}(\vec{r})=-\frac{\alpha_s}{4}{\bf \lambda_i\cdot\lambda_j}\left[\frac{1}{4M_iM_j}\frac{1}{r^3}\right]\hat{S}_{ij}
\end{equation}
where,
\begin{equation}
\hat{S}_{ij}=[3(\vec{\sigma}_i\cdot\hat{r})(\vec{\sigma}_j\cdot\hat{r})-\vec{\sigma}_i\cdot\vec{\sigma}_j]
\end{equation}
The tensor potential is a scalar which is obtained by contracting two second rank tensors. Here, $\hat{r}=\hat{r}_i-\hat{r}_j$ is the unit vector in the direction of $\vec{r}$. In the presence of the tensor interaction, $\vec{L}$ is no longer a good quantum number.\\

The only short-range force between the quarks is spin dependent which comes from OGEP. The spin-orbit couplings give rise to long range forces according to lattice QCD, since the interaction energy between two distant static quarks in lattice QCD involves a power-series expansion in inverse powers of the gluon quark coupling constant\cite{DG75}.  In QCD the Fermi-Breit corrections are large for mesons and baryons. For attractive potentials for the Dirac delta  function $\delta(r)$, the wave equations have no physically acceptable solutions and lead to collapse both for  the quark-antiquark and three quark system. Hence, there are models which have introduced a cut-off or smearing function to weaken the singularity. Though, the finite range function is calculable, it has to be parameterized\cite{IR92}. But it has been shown that the short range spin interactions itself is strong enough to support a bound state of a single $q\bar{q}$ pair. The  tensor and spin-orbit potentials (eqs 7 and 8) are of long range, in contrast to the “zero range” $\delta(r)\sigma_1\cdot\sigma_2$ central potential (eq. 6)\cite{SN86}. But the actual range of the $\delta(r)\sigma_1\cdot\sigma_2$  is still unkown and hence is required to do a non-perturbative calculation ( or a larger basis)\cite{BR80}. Also there are other effects like suppression of the spin orbit potential from OGEP from the spin orbit potential arising from  the confinement potential (they have opposite sign).

\subsection{Radiative Decays}  
We consider two types of radiative transitions of the $B_c$ meson:
a) Electric dipole (E1) transitions are those transitions  in which the orbital quantum number is changed ($\Delta L = 1$,
$\Delta S = 0$). Examples of such transitions are $n ^3S_1\rightarrow n' {^3P_J}\gamma (n > n' )$ and  $n ^3P_J\rightarrow n' {^3S_1}\gamma (n \geq n' )$. The strength of the electric dipole transitions is governed by the size of the radiator and the charges of the constituent quarks. The E1 partial decay width is given by \cite{WJ88},

\begin{equation}
\Gamma_{a\rightarrow b}=\frac{4\alpha}{9}\mu^2\left(\frac{Q_c}{m_c}-\frac{Q_{\bar{b}}}{m_{\bar{b}}}\right)^2\frac{E_b(k_0)}{m_a}k^3_0\left|\bra{R_b}r\ket{R_a}\right|^2
\left\{\begin{array}{cl}
(2J+1)/3,& ^3S_1\rightarrow ^3P_J\\
1/3,&^3P_J\rightarrow ^3S_1\\
1/3,& ^1P_1\rightarrow  ^1S_0\\
1,&^1S_0\rightarrow ^1P_1\end{array}\right.
\end{equation}
where $k_0$ is the energy of the emitted photon,

~~~~~~~~~$k_0=m_a-m_b$ in non relativistic limit.

$\alpha$ is the fine structure constant. $Q_c=2/3$ is the charge of the c quark and $Q_{\bar{b}}=1/3$ is the charge of the $\bar{b}$ quark in units of $|e|$, $\mu$ is reduced mass
$
\mu=\frac{m_bm_c}{m_b+m_c}
$
and\\
~~~~~~~ $\frac{E_b(k_0)}{m_a}=1$ in non relativistic limit.
 \begin{equation}
 \bra{R_b}r\ket{R_a}=\int^\infty_0 r^3R_b(r)R_a(r)dr
 \end{equation}
 is the radial overlap integral which has the dimension of length, with $R_{a,b}(r)$ being the normalized radial wave functions for the corresponding states.\\
 
b) Magnetic dipole (M1) transitions are those transitions in which the spin of the quarks is changed ($\Delta S=1, ~~\Delta L=0$) and thus the initial and final states belong to the same orbital excitation but have different spins. Examples of such transitions are vector to pseudo scalar ($n~^3S_1\rightarrow n'~^1S_0+\gamma$, $n\geq n'$) and pseudo scalar to vector ($n~^1S_0\rightarrow n'~^3S_1+\gamma$, $n>n'$ meson decays.\\

The M1 partial decay width between S wave states is \cite{WJ88,NL78}

\begin{equation}
\begin{split}
\Gamma=\delta_{L_aL_b}4\alpha k^3_0\frac{E_b(k_0)}{m_a}\left(\frac{Q_c}{m_c}+(-1)^{S_a+S_b}\frac{Q_b}{m_b}\right)^2(2S_a+1)
\times(2S_b+1)(2J_b+1)\\
\left \{\begin{array}{ccc}
S_a & L_a & J_a \\
J_b & 1  & S_b 
 \end{array} \right\}^2\left \{\begin{array}{ccc}
1 & \frac{1}{2} & \frac{1}{2} \\
\frac{1}{2} & S_a  & S_b 
 \end{array} \right\}^2
\times\left[\int^\infty_0  R_{n_bL_b}(r)r^2R_{n_aL_a}(r) dr\right]^2
\end{split}
\end{equation} 
which can be further simplified to

\begin{equation}
\Gamma_{M1}(a\to b+\gamma)=\frac{16}{3}\alpha\mu^2_{eff}k^3_0(2J_b+1)\\
\left[\int^\infty_0  R_{n_bL_b}(r)r^2R_{n_aL_a}(r) dr\right]^2
\end{equation} 

where $\int^\infty_0 dr  R_{n_bL_b}(r)r^2R_{n_aL_a}(r)$ is the overlap integral for unit operator between the coordinate wave functions of the initial and the final meson states, $m_c$ and $m_b$ are the masses of the charm and bottom quarks and $\mu^2_{eff}=\frac{m_bQ_c-m_cQ_{\bar{b}}}{4m_cm_b}$. $S_a$, $S_b$, $L_a$, $J_a$ and $J_b$ are the spin quantum number, orbital angular momentum and total angular momentum quantum numbers of initial and final meson states respectively.  \\
The M1 transitions contribute little to the total widths of the 2S levels, since it cannot decay by annihilation.
Allowed M1 transitions correspond to triplet-singlet transitions between S-wave states of the same n quantum number, while hindered M1 transitions are either triplet-singlet or singlet-triplet transitions between S-wave states of different n quantum numbers. The allowed M1 transitions are essentially $1~^3S_1\rightarrow 1~^1S_0$ and $2 ~^3S_1\rightarrow 1~ ^1S_0$.

\subsection{Weak Decays}
\label{weak}
The weak decays of bound state of a quark and an anti-quark,  which carries heavy flavour c and b - enable us to probe the validity of the standard
model of elementary particle interactions and determine several parameters of this model. A rough estimate of the $B_c$ weak decay widths can be done by treating the $\bar{b}$-quark and $c$-quark decays independently so that $B_c$ decays can be divided into three classes \cite{AA99,GS91}$\colon$ (i)the $\bar{b}$-quark decay with spectator $c$-quark, (ii) the $c$-quark decay with spectator $\bar{b}$-quark, and (iii) the annihilation $B^+_c\rightarrow l^+\nu_l$ ($c\bar{s},~u\bar{s}$), where $l=e,~\mu,~\tau$.


\section{Results and Discussion}
\label{sec:RD}
\subsection{Mass Spectra}
\begin{table*}
\tbl{\label{mass1}\bf m$_c$ and m$_b$ for various theoretical models (in GeV).}
{\begin{tabular*}{\textwidth}{@{\extracolsep{\fill}}lrrrrrrl@{}}
\hline
Parameter&\multicolumn{1}{c}{Ref.\cite{AM80}}&\multicolumn{1}{c}{Ref. \cite{EJ78}}&\multicolumn{1}{c}{Ref. \cite{WS81}}&\multicolumn{1}{c}{Ref. \cite{CJ77}}&\multicolumn{1}{c}{ Ref.\cite{DR03}}\\
\hline
$m_c$&1.8&1.48 &1.48 &1.48 &1.55 \\

$m_b$&5.174 &5.18 &4.88 &4.88 &4.88 \\
\hline
\end{tabular*}}
\end{table*}
 The quark-anti quark wave functions in terms of oscillator wave functions corresponding to the relative and center of mass coordinates have been expressed here, which are of the form, 
\begin{equation}
\Psi_{nlm}(r,\theta, \phi) = N (\frac{r}{b})^{l}~L_{n}^{l+\frac{1}{2}}(\frac {r^2}{b^2})\exp(-\frac{r^2}{2b^2})Y_{lm}(\theta, \phi)
\end{equation}
where N is the normalising constant given by 
\begin{eqnarray}
{\lvert N \rvert} ^2={\frac{2n!}{b^3 \pi^{1/2}}} \frac{2^{[2(n+l)+1]}}{(2n+2l+1)!}(n+l)!
\end{eqnarray} 
$L_{n}^{l+\frac{1}{2}}$ are the associated Laguerre polynomials. \\

The wave function used in this calculation (eqn.15), is the standard form of the harmonic oscillator wave functions which has been extensively used in earlier works in atomic, and nuclear physics. The wave function is normalized and the normalization constant is given in eqn.16.
The main advantage of using the harmonic oscillator wave function is that it allows the separation of the motion of the center of mass and has been extensively used to classify the spectra of baryons and mesons \cite{FD69,RP71} and  extending to nucleon-nucleon interaction is straight forward \cite{KS89,FA83,KB93}.  If the basic states are the harmonic oscillator wave functions, then it is straightforward to evaluate the matrix elements of few body systems such as mesons or baryons. Since the basic states are the products of the harmonic oscillator wave functions they can be chosen in a manner that allows the product wave functions to be expanded as a finite sum of the corresponding products for any other set of Jacobi coordinates.  
It is advantageous to use the Gaussian form since in the annihilation of quark-anti quark into lepton pairs, the amplitude of the emission or absorption processes depend essentially on the overlap of initial and final hadrons and hence the overlap depends only on the intermediate  distance region of the spatial wave functions  which can extend up to 0.5 fm. This intermediate region can be described by potentials that are similar in this region and hence harmonic oscillator wave functions are expected to reproduce emission and absorption processes quite well.\\
 
The four parameters in our model are the mass of charm quark $m_c$, the mass of bottom quark $m_b$, the harmonic oscillator size parameter $b$ and the quark-gluon coupling constant $\alpha_{s}$. There are several papers in literature where the size parameter $b$ is defined \cite{SN86,IM92}. The value of $b$ is fixed by minimizing the expectation value of the Hamiltonian for the vector meson. The confinement strength $a_c$ is fixed by the stability condition for variation of mass of the vector meson against the size parameter $b$. To fit $\alpha_s$ , $m_b$ and $m_c$, we start with a set of reasonable values and diagonalize the matrix for $B_c$ meson. Then we tune these parameters to obtain an agreement with the experimental value for the mass of $B_c$ meson. In literature we find different sets of values for $m_c$ and $m_b$, which are listed in table \ref{mass1}.

The values of strong coupling constant $\alpha_s$ in literature are listed in table \ref{alpha}. The value of strong coupling constant ($\alpha_{s}$=0.3) used is compatible with the perturbative treatment.\\
We use the following set of parameter values.
\begin{equation}
\begin{split}
m_c = 1480.0~~~{\rm MeV};~~~m_b=4750.0~{\rm MeV};\\
b= 0.350~{\rm fm};~~~ \alpha_s = 0.300;~~a_c=145~{\rm MeV~fm^{-1}};
\end{split}
\end{equation}
\begin{table*}[!ht]
\tbl{\bf $\alpha_s$ for various theoretical models.}
{\begin{tabular*}{\textwidth}{@{\extracolsep{\fill}}lrrrrrrl@{}}
\hline
Parameter&\multicolumn{1}{c}{Ref. \cite{SN85}}&\multicolumn{1}{c}{Ref. \cite{DR03}}&\multicolumn{1}{c}{Ref.\cite{AA05}}&\multicolumn{1}{c}{ Ref. \cite{EC94}}&\multicolumn{1}{c}{ Ref. \cite{SA95}}\\
\hline
$\alpha_s$&0.21&0.265&0.357&0.361&0.391\\
\hline
\end{tabular*}\label{alpha}}
\end{table*}
 
 The mass spectrum has been obtained by diagonalizing the Hamiltonian in a large basis of $15\times 15$ matrix which has not been carried out in other existing models which is a new ingredient in our model. The calculation clearly indicates that masses for both pseudo scalar and vector mesons converge to the experimental values when the diagonalization is carried out in a larger basis. In our earlier work also, we had come to the similar conclusion while investigating light meson spectrum\cite{VH04,BK05}. The diagonalization of the Hamiltonian matrix in a larger basis leads to the lowering of the masses and justifies the perturbative technique to calculate the mass spectrum. The calculation clearly indicates that when diagonalization is carried out in a larger basis convergence is achieved both for pseudo-scalar mesons and vector mesons to the respective experimental values. \\
 
  For the case of a bound system of quark and anti-quark of unequal mass, charge conjugation parity is no longer a good quantum number so that the states with different total spins but with the same total angular momentum, such as the $^3P_1 - ^1P_1$ and $^3D_2 - ^1D_2$ pairs, can mix via the spin orbit interaction or some other mechanism. The $B_c$ meson states with $J=L$ are linear combination of spin triplet $\ket{^3L_J}$ and spin singlet $\ket{^1L_J}$ states which we describe by the following mixing$\colon$
 \begin{eqnarray}
&\ket{nL'}=\ket{n~^1L_J}\cos\theta_{nL}+\ket{n~^3L_J}\sin\theta_{nL}\\
&\ket{nL}=-\ket{n~^1L_J}\sin\theta_{nL}+\ket{n~^3L_J}\cos\theta_{nL}
\end{eqnarray}
$$~J=L=1,2,3,\cdots$$
\begin{longtable}[!ht]{@{\extracolsep{\fill}}cccccccccccccc}
\caption{\label{spectrum}\bf $B_c$ meson mass spectrum (in GeV).}\\
\hline
State &&\\
$n~^{2S+1}L_J$&\multicolumn{1}{c}{This work}&\multicolumn{1}{c}{Ref.\cite{SJ96}}&\multicolumn{1}{c}{Ref. \cite{VA95}}&\multicolumn{1}{c}{Ref. \cite{ZV95}}&\multicolumn{1}{c}{Ref. \cite{EC94}}&\multicolumn{1}{c}{Ref.\cite{DR03}}&\multicolumn{1}{c}{Ref.\cite{SN85}}&\multicolumn{1}{c}{Ref.\cite{CT96}}&\multicolumn{1}{c}{Ref.\cite{FL}}\\
\hline
$1~^1S_0$&6.275&6.247&6.253&6.260&6.264&6.270&6.271&6.280$\pm 30\pm 190$&6.286\\

$1~^3S_1$&6.357&6.308&6.317&6.340&6.337&6.332&6.338&6.321$\pm 20$&6.341\\
$1~^3P_0$&6.638&6.689&6.683&6.680&6.700&6.699&6.706&6.727$\pm 30$&6.701\\
$1P$&6.686&6.738&6.717&6.730&6.730&6.734&6.741&6.743$\pm 30$&6.737\\
$1P'$&6.734&6.757&6.729&6.740&6.736&6.749&6.750&6.765$\pm 30$&6.760\\
$1~^3P_2$&6.737&6.773&6.743&6.760&6.747&6.762&6.768&6.783$\pm 30$&6.772\\
$2~^1S_0$&6.862&6.853&6.867&6.850&6.856&6.835&6.855&6.960$\pm 80\pm$&6.882\\
$2~^3S_1$&6.897&6.886&6.902&6.900&6.899&6.881&6.887&6.990$\pm 80$&6.914\\
$1~^3D_1$&6.973&&7.008&7.010&7.012&7.072&7.028&&7.019\\
$1D$&6.974&&7.001&7.020&7.012&7.077&7.041&&7.028\\
$1D'$&7.003&&7.016&7.030&7.009&7.079&7.036&&7.028\\
$1~^3D_3$&7.004&&7.007&7.040&7.005&7.081&7.045&&7.032\\
$2~^3P_0$&7.084&&7.088&7.100&7.108&7.091&7.122&&\\
$2P$&7.137&&7.113&7.140&7.135&7.126&7.145&&\\
$2P'$&7.173&&7.124&7.150&7.142&7.145&7.150&&\\
$2~^3P_2$&7.175&&7.134&7.160&7.153&7.156&7.164&&\\
$3~^1S_0$&7.308&& &7.240&7.244&7.193&7.250&&\\
$3~^3S_1$&7.333&& &7.280&7.280&7.235&7.272&&\\
$2~^3D_1$&7.377&&\\
$2D$&7.385&& \\
$2D'$&7.408&&\\
$2~^3D_3$&7.410&&\\
$3~^3P_0$&7.492&&\\
$3P$&7.546&&\\
$3P'$&7.572&&\\
$3~^3P_2$&7.575&&\\
$4~^1S_0$&7.713&& &&  7.562\\
$4~^3S_1$&7.734&& &&  7.594\\
$3~^3D_1$&7.761&&\\
$3D$&7.781&& \\
$3D'$&7.783&&\\
$3~^3D_3$&7.796&&\\
$4~^3P_0$&7.970&&\\
$4P$&7.943&&\\
$4P'$&7.942&&\\
$4~^3P_2$&7.970&&\\
$5~^1S_0$&8.097&& \\
$5~^3S_1$&8.115&& \\
\hline
\caption{\label{spectrum}: (continued)}\\
\hline
State &&\\
$n~^{2S+1}L_J$&This work&&$n~^{2S+1}L_J$&This work&&$n~^{2S+1}L_J$&This work\\
\hline
$4~^3D_1$&8.132&&$9~^1S_0$&9.543&&$12P'$&10.778\\
$4D$&8.154&& $9~^3S_1$&9.553&&$12~^3P_2$&10.787\\
$4D'$&8.155&&$8~^3D_1$&9.552&&$13~^1S_0$&11.015\\
$4~^3D_3$&8.168&&$8D$&9.573&&$13~^3S_1$&11.020\\
$5~^3P_0$&8.254&&$8D'$&9.572&&$12~^3D_1$&10.964\\
$5P$&8.314&&$8~^3D_3$&9.586&&$12D$&10.972 \\
$5P'$&8.312&& $9~^3P_0$&9.695&&$12D'$&10.971\\
$5~^3P_2$&8.337&&$9P$&9.646&&$12~^3D_3$&10.977\\
$6~^1S_0$&8.469&&$9P'$&9.731&&$13~^3P_0$&11.192\\
$6~^3S_1$&8.484&&$9~^3P_2$&9.749&&$13P$&11.016\\
$5~^3D_1$&8.494&&$10~^1S_0$&9.896&&$13P'$&11.203\\
$5D$&8.516&& $10~^3S_1$&9.905&&$13~^3P_2$&11.208\\
$5D'$&8.517&&$9~^3D_1$&9.893&&$14~^1S_0$&11.458\\
$5~^3D_3$&8.530&&$9D$&9.912&&$14~^3S_1$&11.462\\
$6~^3P_0$&8.621&&$9D'$&9.912&&$13~^3D_1$&11.356\\
$6P$&8.675&&$9~^3D_3$&9.925&&$13D$&11.361 \\
$6P'$&8.674&& $10~^3P_0$&10.051&&$13D'$&11.361\\
$6~^3P_2$&8.696&&$10P$&9.855&&$13~^3D_3$&11.365\\
$7~^1S_0$&8.832&&$10P'$&10.081&&$14~^3P_0$&11.623\\
$7~^3S_1$&8.846&&$10~^3P_2$&10.097&&$14P$&11.544\\
$6~^3D_1$&8.851&&$11~^1S_0$&10.253&&$14P'$&11.629\\
$6D$&8.872&& $11~^3S_1$&10.261&&$14~^3P_2$&11.632\\
$6D'$&8.873&&$10~^3D_1$&10.228&&$15~^1S_0$&12.005\\
$6~^3D_3$&8.886&&$10D$&10.243&& $15~^3S_1$&12.008\\
$7~^3P_0$&8.981&&$10D'$&10.242&&$14~^3D_1$&11.801\\
$7P$&9.027&&$10~^3D_3$&10.253&&$14D$&11.804 \\
$7P'$&9.029&&$11~^3P_0$&10.403&&$14D'$&11.804\\
$7~^3P_2$&9.051&&$11P$&10.172&&$14~^3D_3$&11.806\\
$8~^1S_0$&9.189&&$11P'$&10.428&&$15~^3P_0$&12.175\\
$8~^3S_1$&9.201&&$11~^3P_2$&10.441&&$15P$&11.917\\
$7~^3D_1$&9.203&&$12~^1S_0$&10.621&&$15P'$&12.179\\
$7D$&9.224&& $12~^3S_1$&10.628&&$15~^3P_2$&12.180\\
$7D'$&9.224&&$11~^3D_1$&10.604&&$15~^3D_1$&12.350\\
$7~^3D_3$&9.238&&$11D$&10.617&&$15D$&12.351 \\
$8~^3P_0$&9.339&&$11D'$&10.616&&$15D'$&12.351\\
$8P$&9.366&&$11~^3D_3$&10.625&&$15~^3D_3$&12.352\\
$8P'$&9.381&&$12~^3P_0$&10.762&&\\
$8~^3P_2$&9.401&&$12P$&10.561&&\\

\end{longtable}%
where $\theta_{nL}$ is a mixing angle, and the primed state has the heavier mass.  The values of the mixing angle for P states are $\theta_{1P}=0.2^\circ$, $\theta_{2P}=0.10^\circ$ and $\theta_{3P}=0.05^\circ$\\

Similarly for $L=J=2$ we have mixing of D states and the values of mixing angles for D states are
$\theta_{1D}=0.20^\circ$ and $\theta_{2D}=0.05^\circ$.\\
 The calculated masses of the $c\bar{b}$ states after diagonalization are listed in Table ~\ref{spectrum}. Our calculated mass value for $B_c(1S)$ is 6275.75 MeV which agrees withe experimental value 6.275 GeV\cite{PDG} and for $B^*_c$(1S) is 6357.27 MeV. $B^*_c$(1S) is heavier than $B_c$(1S) by 81.52 MeV. This difference is justified by calculating the $^3S_1-{}^1S_0$ splitting of the ground state which is given by 
 \begin{equation}
 M({}^3S_1)-M({}^1S_0)=\frac{32\pi\alpha_s|\psi(0)|^2}{9m_cm_b}
 \end{equation}
 The mass of first radial excitation $B_c$(2S) is 6862.88 MeV which is heavier than $B_c$(1S) by 587.13 MeV. This value agrees with the experimental value of $B_c$(2S) 6842$\pm$4$\pm$5 \cite{AG14}. The difference between the $B^*_c$(2S) and $B^*_c$(1S) masses turns out to be 540.14 MeV.  Our prediction for masses of orbitally excited $c\bar{b}$ states are in good agreement with the other model calculations. Some of the states (i.e., $2^3P_0$, 2P1,$2P1'$,$2^3P_2$) are 50-100 MeV heavier in our model.\\

Vijande {\textit{et al.}} have derived Coulomb strength $A_{3Q}\approx \frac{1}{2}A_{Q\bar{Q}}$ and confinement strength $B_{3Q}\approx B_{Q\bar{Q}}$  from the triply baryon spectra calculated in Lattice QCD \cite{VJ14,VJ15,VJ16}. They obtained a nice fit of the nonperturbative QCD results with the Cornell like potential 
\begin{equation}
V^{3Q}(r)=-A\sum_{i<j}\frac{1}{|\vec{r}_i-\vec{r}_j|}+B\sum_{i<j}|+C
\end{equation}
 with Coulomb strength $A_{3Q}=0.1875$ and confinement strength $B_{3Q}=0.1374~ \rm{GeV}^2$. The Coulomb strength and the confinement strength calculated in our model is $A_{Q\bar{Q}}=0.4$ and $B_{Q\bar{Q}}=0.1526$ GeV$^2$ (i.e.  coefficients of Coulomb and linear potentials) respectively. Using these values we come to the conclusion that $\frac{A_{3Q}}{A_{Q\bar{Q}}}\approx \frac{1}{2}$ and $B_{3Q}\approx B_{Q\bar{Q}}$. Vijande {\textit{et al.}} obtained $\frac{A_{3Q}}{A_{Q\bar{Q}}}<\frac{1}{2}$ slightly different from $\frac{1}{2}$ as the one gluon exchange result and our result \cite{VJ15}.

\subsection{Radiative Decays}

The calculation of radiative (EM) transitions between the meson states can be performed from first principles in lattice QCD, but these calculation techniques are still in their development stage. At present, the potential model approaches provide the detailed predictions that can be compared to experimental results. In our non relativistic model we consider the Magnetic dipole (M1) transitions and Electric dipole (E1) transitions of $B_c$ meson.\\

We have listed the possible $E1$ decay modes in table \ref{E1} and given the predictions for E1 decay widths. Also we have compared our predictions with other theoretical models.
  Most of the predictions for $E1$ transitions are in qualitative agreement. However, there are some differences in the predictions  due to differences in phase space arising from different mass predictions and also from the wave function effects. For the transitions involving $P1$ and $P1'$ states which are mixtures of the spin singlet $^1P_1$ and spin triplet $^3P_1$ states, there exists huge difference between the different theoretical predictions. These may be due to the different $^3P_1 - ^1P_1$ mixing angles predicted by the different models. Wave function effects also appear in decays from radially excited states to ground state mesons such as $2~^3P_0\rightarrow 1~^3S_1\gamma$. The overlap integral for these transitions in our model vanishes and hence we get decay width for these transitions zero. This is due to the orthogonality condition for wave functions.\\
  
 The possible radiative M1 transition modes are (1) 2 $^3S_1\rightarrow 2 ^1S_0+\gamma$, (2) 2 $^3S_1\rightarrow 1 ^1S_0+\gamma$, (3) 2 $^1S_0\rightarrow 1 ^3S_1+\gamma$ and (4)1 $^3S_1\rightarrow 1 ^1S_0+\gamma$. 

In the above (2) and (3) represent hindered transitions and (1) and (4) represent allowed transitions. 
 
The resulting M1 radiative transition rates of these states are presented in table \ref{M1}. In this table we give the calculated M1 decay widths for allowed transitions (n$^3S_1 \rightarrow n' {^1S_0} + \gamma$,
$n = n'$ ) and we compare the decay widths with other non relativistic quark models \cite{EC94,SA95,FL}.  The hindered transitions are strongly suppressed in the non relativistic limit due to the orthogonality of the initial and final state wave functions. By adding relativistic effects to the wave function the hindered M1 transition rates can be enhanced.

\subsection{Weak Decay and Life Time of $B_c$ meson} 
 \begin{table*}[!h]
\tbl{\label{E1}\bf E1 transition rates of $B_c$ meson.}
{\begin{tabular*}{\textwidth}{@{\extracolsep{\fill}}lrrrrrrrrrrl@{}}
\hline
Transition&\multicolumn{1}{c}{k$_0$}&\multicolumn{1}{c}{$\Gamma(i\rightarrow f+\gamma)$}&\multicolumn{1}{c}{Ref. \cite{DR03}}&\multicolumn{1}{c}{Ref. \cite{EC94}}&\multicolumn{1}{c}{Ref. \cite{VA95}}&\multicolumn{1}{c}{Ref.\cite{FL}}\\
&MeV&keV&keV&keV&keV&keV\\
\hline
$1^3P_0\rightarrow 1^3S_1\gamma$&281.22&30.67&75.5&79.2&65.3&74.2\\
$1P\rightarrow 1^3S_1\gamma$&329.71&49.438&87.1&99.5&77.8&75.8\\
$1P'\rightarrow 1^3S_1\gamma$&377.72&74.331&13.7&0.1&8.1&26.2\\
$1^3P_2\rightarrow 1^3S_1\gamma$&380.55&112.754&122&112.6&102.9&126\\
$1P\rightarrow 1^1S_0\gamma$&411.23&31.974&18.4&0&11.6&32.5\\
$1P'\rightarrow 1^1S_0\gamma$&459.24&44.531&147&56.4&131.1&128\\
$2^3S_1\rightarrow 1^3P_0\gamma$&258.92&7.98&5.53&7.8&7.7&9.6\\
$2^3S_1\rightarrow 1P\gamma$&210.43&8.568&7.65&14.5&12.8&13.3\\
$2^3S_1\rightarrow 1P'\gamma$&162.42&3.939&0.74&0&1.0&2.5\\
$2^3S_1\rightarrow 1^3P_2\gamma$&159.59&6.228&7.59&17.7&14.8&14.5\\
$2^1S_0\rightarrow 1P\gamma$&175.9&5.004&1.05&0&1.9&6.4\\
$2^1S_0\rightarrow 1P'\gamma$&127.89&1.923&4.40&5.2&15.9&13.1\\
$2^3P_0\rightarrow 1^3S_1\gamma$&726.95&0&&21.9&16.1\\
$2P\rightarrow 1^3S_1\gamma$&797.33&0&&22.1&15.3\\
$2P'\rightarrow 1^3S_1\gamma$&816.04&0&&2.1&2.5\\
$2^3P_2\rightarrow 1^3S_1\gamma$&818.37&0&&25.8&19.2\\
$2P\rightarrow 1^1S_0\gamma$&861.56&0&& &3.1\\
$2P'\rightarrow 1^1S_0\gamma$&897.56&0&& &20.1\\
$2^3P_0\rightarrow 2^3S_1\gamma$&186.81&14.987&34.0&41.2&25.5\\

$2P\rightarrow 2^3S_1\gamma$&239.9&31.739&45.3&54.3&32.1\\
$2P'\rightarrow 2^3S_1\gamma$&275.9&48.280&10.4&5.4&5.9\\
$2^3P_2\rightarrow 2^3S_1\gamma$&278.23&49.513&75.3&73.8&49.4\\
$2P\rightarrow 2^1S_0\gamma$&274.43&47.512&13.8& &8.1\\
$2P'\rightarrow 2^1S_0\gamma$&310.43&68.770&90.5&&58.0\\
\hline
\end{tabular*}}
\end{table*}

\begin{table*}[!h]
\tbl{\label{M1}\bf M1 transition rates for the $B_c$ meson.}
{\begin{tabular*}{\textwidth}{@{\extracolsep{\fill}}lrrrrrl@{}}
\hline
Transition&\multicolumn{1}{c}{$k_{0}$}&\multicolumn{1}{c}{Ref. \cite{FL}}&\multicolumn{1}{c}{Ref. \cite{SA95}  }&\multicolumn{1}{c}{Ref. \cite{DR03}}&\multicolumn{1}{c}{Ref.\cite{EC94}}&\multicolumn{1}{c}{This work}\\
&$\Gamma(keV)$&$\Gamma(keV)$&$\Gamma(keV)$&$\Gamma(keV)$&$\Gamma(keV)$&$\Gamma(keV)$\\
\hline
$1~ ^3S_1\rightarrow 1 ^1S_0\gamma$&81.52&0.190&0.060&0.073&0.135&0.0581\\
$2~ ^3S_1\rightarrow 2 ^1S_0\gamma$&34.53&0.043&0.010&0.030&0.029&0.00173\\
\hline
\end{tabular*}}
\end{table*}

\begin{table*}[!h]
\centering
\caption{\label{T1}\bf Comparison of life time of $B_c$ meson (in ps).}
\begin{tabular*}{\textwidth}{@{\extracolsep{\fill}}lrrrrrrl@{}}
\hline
This work&\multicolumn{1}{c}{Experiment\cite{PDG}}&\multicolumn{1}{c}{Ref.\cite{AA99}}&\multicolumn{1}{c}{Ref.\cite{VA95} }&\multicolumn{1}{c}{Ref.\cite{KVV}}&\multicolumn{1}{c}{Ref. \cite{GS85}}\\
\hline
0.379&0.452$\pm 0.033$&0.47&0.55$\pm 0.15$&0.50&0.75\\
\hline
\end{tabular*}
\end{table*} 

\begin{table*}[!h]
\centering
\caption{\label{fbc}\bf Comparison of predictions for the pseudo scalar decay constant of the $B_c$ meson.}
\begin{tabular*}{\textwidth}{@{\extracolsep{\fill}}lrrrrrrl@{}}
\hline
Parameter&\multicolumn{1}{c}{Ref.\cite{WS81}}&\multicolumn{1}{c}{Ref. \cite{AM80}}&\multicolumn{1}{c}{Ref.\cite{CJ77}}&\multicolumn{1}{c}{Ref.\cite{CT96}}&\multicolumn{1}{c}{This work}\\
\hline
$f_{B_c}$&500&512&479&440$\pm$20&439.735\\
\hline
\end{tabular*}
\end{table*}
In accordance with the classification given in section \ref{weak}, the total decay width can be written as the sum over partial widths
 \begin{equation}
 \Gamma(B_c\rightarrow X)=\Gamma_1(\bar{b}\rightarrow X)+\Gamma_2(c\rightarrow X)+\Gamma_3(ann)
 \end{equation}
 In the spectator approximation:
 \begin{equation}
 \Gamma_1(\bar{b}\rightarrow X)=\frac{9G^2_F|V_{cb}|^2m^5_b}{192\pi^3}\label{eq1}
 \end{equation}
 Calculated value of $\Gamma_1(\bar{b}\rightarrow X)$ is $9.628\times 10^{-4}~\rm{eV}$ 
 and
\begin{equation}
 \Gamma_2(c\rightarrow X)=\frac{5G^2_F|V_{cs}|^2m^5_c}{192\pi^3} \label{eq2}
 \end{equation}
 where $V_{cb}$ and $V_{cs}$ are the elements of the CKM matrix. The decay widths are calculated using $|V_{bc}|=0.044$ \cite{PDG} and $|V_{cs}|=0.975$ \cite{PDG}.\\
 Calculated value of $\Gamma_2(c\rightarrow X)$ is $7.712\times 10^{-4}~\rm{eV}$.\\
 
The decay of vector meson into charged leptons proceeds through the virtual photon $(q\bar{q}\rightarrow l^+l^-)$. The $^3S_1$ and $^3D_1$ states have quantum numbers of a virtual photon, $J^{PC}=1^{--}$ and can annihilate into lepton pairs through one photon.
Annihilation widths such as $c\bar{b}\rightarrow l\nu_l$ are given by the expression
\begin{equation}
\Gamma_3(ann)=\frac{G^2_F}{8\pi}|V_{bc}|^2f^2_{B_c}M_{B_c}\sum_i m^2_i\left(1-\frac{m^2_i}{M^2_{B_c}}\right)C_i\label{eq3}
\end{equation}
where $m_i$ is the mass of the heavier fermion in the given decay channel. For lepton channels $C_i=1$ while for quark channels $C_i=3|V_{q\bar{q}}|^2$. \\
Calculated value of $\Gamma_3$ is $3.56\times 10^{-6}~\rm{eV}$\\
Adding these results we get the total decay width~~~~\\ $\Gamma(\rm{total})=\Gamma_1+\Gamma_2+\Gamma_3=18.104\times 10^{-4}~\rm{eV}$ corresponding to a life time of $\tau=0.364~\rm{ps}$.\\

The pseudo scalar decay constant $f_{B_c}$ is defined by \cite{EC94}
\begin{equation}
\bra{0}\bar{b}(x)\gamma^\mu\gamma_5 c(x)\ket{B_c(k)}=if_{B_c}V_{cb}k^\mu 
\end{equation}  
 where $k^\mu$ is the four-momentum of the $B_c$ meson.
 In the non relativistic limit the pseudo scalar decay constant is proportional to the wave function at the origin and is given by van Royen-Weisskopf formula \cite{RV67}
 \begin{equation}
 f_{B_c}=\sqrt{\frac{12}{M_{B_c}}}\psi(0)
 \end{equation}
The value of decay constant in the non relativistic potential model is listed in table \ref{fbc}.

\section{Theoretical uncertainties in the predictions of the model}
The goal of the present work is to obtain a reliable estimate of the masses of the $c\bar{b}$ states and the decay widths and understand the uncertainties in the calculation in the frame work of non-relativistic quark models. The main reason for using the non-relativistic model is, it gives a good account of the spectra and it is possible to separate the center-of-mass motion and the relativistic correction can be incorporated by $v/c$ expansion.  For $c\bar{b}$  system quark velocities are sizeable and hence needs significant corrections, but are still very small to make any significant changes in the spectra or decay widths and $v/c$ corrections are very small and the non-relativistic predictions radiative decays are very accurate\cite{LE88}.  The non-relativistic quark models have many other basic features of the QCD and has met with great success in predicting the spectra and decay widths of hadrons as it allows direct calculations of the relevant matrix elements for each hadron.

  The standard way of estimating the uncertainties in any model is to vary different parameters in the model. 

Also it is known that using a larger harmonic oscillator basis increases the wave function at the origin since the higher order states mixing into the wave function can probe the short distance of the potential\cite{IS89,BK05}. Hence there is a slight theoretical uncertainty in the value of the oscillator size parameter (b). Also, there is theoretical uncertainty in the form of the wave function $\psi(0)$ at the origin. The $\psi(0)$ is relatively flat for linear and harmonic oscillator potentials, but it raises sharply for the Columbic potential. Ultimately, the form of the wave function at the origin has to be settled by lattice QCD calculations which will be the most reliable results as they are calculations from the first principles. But, for the orbitally excited states, the wave function at the origin vanishes in the non relativistic limit and hence the lowest order approximation vanishes.  In calculating the masses and decay widths of quarkonium there are ambiguities in the value of $\alpha_s$. These ambiguities are both theoretical and experimental. In theory there is uncertainty in the scale $\mu$ to be considered in computing $\alpha_s(\mu^2)$.  On the other hand in experimental side, there is uncertainty in the scale parameter $\Lambda$ of the QCD. 

Another source of theoretical uncertainty is the masses of the charm ($m_c$) and bottom quark ($m_b$).  For the charm quark, masses of $m_c$ used in literature are 1 GeV, 1.5 GeV and 1.8 GeV. But it should be noted that the spectra of the quarkonium are not very sensitive to the mass of the quark, but $|\psi(0)|^2$ is strongly dependent on $m_c$. But, since we are making use of the non-relativistic model both for spectrum and decay the choices for the masses of $m_c$ are minimum. The $m_c$ is constrained by the rate of the M1 transition $J/\psi\to\eta_c\gamma$. This branching ratio restricts $m_c\sim$ 2 GeV. Our choice for $m_c$ is 1.48 GeV which is the value quoted in PDG.  For the b quark, the mass is again taken from PDG. For the b quark, for a fixed energy eigen value the dependence on $|\psi(0)|^2$ is minimal as the variation is little because of smaller uncertainty of the $m_b$. 
\section{Conclusions}
\label{sec:C}
The study highlights the mass spectra of $c\bar{b}$ meson in a non relativistic quark model. The ground state mass of $c\bar{b}$ state  calculated in our model matches the experimental data. When the results for $c\bar{b}$ state mass spectrum are compared with the previous calculations, it is found that the predictions for the mass spectrum agree within a few MeV. The differences between the predictions in most cases do not exceed 30 MeV and the higher orbitally excited states are 50-80 MeV heavier in our model. The hyperfine splitting of the ground state  vector and pseudo scalar $c\bar{b}$ states in our model is in good agreement with the prediction made by Penin et al. They predicted the hyperfine splitting of the vector and pseudo scalar $B_c(1S)$ mesons to be $M(B^*_c)-M(B_c)=50\pm 17(th) \rm{MeV}$. The ground state pseudo scalar $B_c$ and vector $B^*_c$ meson masses lie within the ranges quoted by Kwong and Rosner in their survey of techniques for estimating the masses of the $c\bar{b}$ ground state: i.e., $6.194~ \rm{GeV}<M_{B_c}<6.292~\rm{GeV}$ and $6.284~\rm{GeV}<M_{B^*_c}<6.357~\rm{GeV}$. \\

 Radiative decays are the dominant decay modes of the $c\bar{b}$ excited states having widths of about a fraction of MeV. In order to understand the $c\bar{b}$ spectrum and distinguishing exotic states, it is very essential that the masses and the radiative decay widths of $c\bar{b}$ states are accurately determined. The calculated M1 transition rates reasonably agree with the other theoretical model predictions as listed in table \ref{M1}. It is clearly seen in this calculation that the relativistic effects play an important role in determining the M1 radiative transition rates, since the hindered transition rates are zero due to the wave function orthogonality in the NRQM formalism. The inclusion of these relativistic effects may enhance the hindered transition rates and reduce the allowed transition rates.\\
 
  Most of our predictions for the E1 decay rates are in good agreement with the other theoretical calculations. The differences in the prediction for the decay rates in various theoretical models can be attributed to differences in mass predictions, wave function effects and singlet - triplet mixing angels. We have done an estimation of weak decay widths in the spectator quark approximation and calculated the life time of $c\bar{b}$ state. We get about 53\% branching ratio for $b$-quark decays, about 42\% for $c$-quark decays and about 5\% branching ratio in annihilation channel. The life time of $B_c$ meson predicted in in othe theoreticla model is listed in table \ref{T1}. The life time of $c\bar{b}$ state predicted in this calculation is found to be in good agreement with experimental value as well as with other theoretical predictions. The decay constant of $c\bar{b}$ state ($f_{B_c}$) has been calculated and compared with other model predictions and it is found that the decay constant is consistent with these predictions.  \\

 A simple non relativistic model employing OGEP and linear confinement potential used in this study is successful to predict the various properties of $c\bar{b}$ states and this can shed further light on their non leptonic transition rates.  
\begin{center}
\textbf{Acknowledgements}
\end{center}
One of the authors (APM) is grateful to BRNS, DAE, India for granting the project and JRF (37(3)/14/21/2014BRNS).


\bibliography{mybib}

\end{document}